# Homogeneous Dislocation-Induced Rainbow Concentrating for Elastic Waves

Zi-Dong Zhang[1†], Shi-Li Yang[1†], Shi-Ling Yan[1], Si-Yuan Yu[1,2*], Ming-Hui Lu[1,2*], Yan-Feng Chen[1,2]

[1]National Laboratory of Solid State Microstructures & Department of Materials Science and Engineering, Nanjing University, Nanjing 210093, China

[2]Jiangsu Key Laboratory of Artificial Functional Materials, Nanjing University, Nanjing 210093, China

[†]Zi-Dong Zhang and Shi-Li Yang contributed equally to this work.

[*]Corresponding author. yusiyuan@nju.edu.cn; luminghui@nju.edu.cn

***Abstract***

Defects play a crucial role in the physical properties of crystals, whether for classical or quantum systems. For example, in photonic/phononic crystals, defects can serve as precise guidance and localization of classical electromagnetic or mechanical waves. Rainbow concentrating, a recently proposed exotic wave localization [*Phys. Rev. Lett.* **126**, 113902 (2021)] exploits defects to enable the collection and frequency routing of weak signals in real space. In this paper, using a solid-state phononic crystal (PnC) plate, we experimentally verify this phenomenon by deliberately infusing a homogeneous graded dislocation, *i.e.*, a line defect, into the PnC. Two PnCs separated by the defect will breed deterministic interface states along with the defect, offering rainbow trapping and concentrating for elastic waves. Our PnC-based rainbow trappers and concentrators are scalable and configurable, promising for advancing applications like energy harvesting, information processing, and acoustofluidic manipulating.

## I. Introduction

Accurately guiding and trapping waves, whether in electromagnetics [1-3] or mechanics [4-12], lie at the heart of artificial microstructures and metamaterials and are crucial for realizing applications, *e.g.*, sensing, energy harvesting [13], and information processing [14]. For periodic microstructures such as photonic crystals (PCs) and phononic crystals (PnCs), the introduction of defects into the crystals can bring about defect modes, which can be used to guide and trap those classical waves. Since the frequencies of these defect modes are closely related to the defects themselves, if a crystal has spatially continuous and graded defects, it can be used to distribute waves of different frequencies to different locations in real space, *i.e.*, the rainbow trapping [15]. For electromagnetic waves and airborne sounds, rainbow trapping has been well observed in both one-dimensional (1D) and two-dimensional (2D) systems [16-24]. For a 1D (chain-like) elastic system, rainbow trapping has been achieved by band structure engineering based on chirped PnCs [25], an elastic waveguide loaded with an array of resonators [26-28], a graded piezoelectric metamaterial beam [29]. For the 2D (plate-like) elastic system, however, the development of this effect/function is relatively lagging behind. In a 2D elastic system, despite a variety of proposed rainbow trappers so far, all these solutions require relatively complex technical means (either manipulating the thickness of the plate [30] or the height of the column resonators [31]), which brings great difficulties to sample preparation in continuous elastic media. In 2020, Nakata et al. devise a scheme that produces a series of zero-dimensional localized states in a defect created by a translational deformation of a periodic potential [32]. This scheme allows frequency modulation of defect modes simply by translating the microstructures, undoubtedly providing a convenient and feasible path for the 2D elastic rainbow phenomena. This method has recently been extended theoretically to electromagnetic waves [33]. Lu et al. demonstrated the rainbow trapper and further rainbow concentrator by region-selectively introducing translation to nanostructures in a PC. The so-called rainbow concentrator is an extension of the rainbow trapper. It greatly expands the working direction of the latter, from unidirectional to multi-directional, which is more efficient and conducive to practical applications.

In this paper, based on a 2D PnC plate, we experimentally realized both the elastic rainbow trapping and concentrating. By infusing a homogeneous dislocation, *i.e.*, a line defect, into the PnC, two separated PnCs will have distinct Zak phases, leading to deterministic interface states along with the defect. Different translations can modulate the group velocity of these interface states, offering the verification of rainbow trapping and concentrating for elastic waves. All these PnC-based rainbow trappers and concentrators are convenient to design and configurable. By setting different degrees of defect gradient, both mode volume and frequency spacing of adjacent rainbow modes are adjustable.

## II. Elastic deterministic interface states on a homogeneous dislocation

Our elastic rainbow PnC is designed from a plate with mounted columns in a 2D square lattice, as illustrated in Fig. 1(a). The unit cell of the PnC is a square plate with identical square columns on both sides, placed in the center of the cell. In this study, the PnC is made from aluminum alloy for demonstration. Its lattice constant $a$ is 2 cm; the plate thickness $d$, the height of the square columns $H$, and their lateral size $b$ equal $0.4a$, $0.35a$, and $0.5a$, respectively.

If the columns are mounted on only one side of the plate, the plate's mirror symmetry along its center plane will be broken, leading to ubiquitous hybridization of the extensional and shear horizontal (SH) modes and out-of-plane flexural modes. For the convenience of this study, we chose to have the same columns on both sides of the plate. In this plate PnC, the pure out-of-plane modes are now independent (decoupled) from the others [34]. In most scenarios, the two types of modes cannot be converted into each other so that they can be studied and exploited separately.

Fig. 1(b) shows the calculated band structure of this plate PnC, with the inset corresponding to its Brillouin zone (BZ). By calculating a polarization index, the two types of modes are well distinguished [See Supplementary Information (SI) Note 1 [35]]. In this article, we choose only one of them for experimental demonstration [*i.e.*, the modes with out-of-plane components (specifically, flexural modes)]; in fact, the design scheme for both of them is almost the same. The band structure shows that the flexural modes have a full bandgap from ~58 kHz to ~80 kHz in our plate PnC.

Next, we apply a 2D translation vector $\xi = (\xi_x, \xi_y)$ to all the columns of the PnC, parallel to the plate surface. This translation vector can be regarded as a synthetic dimension, which, together with the Bloch wave vector $k = (k_x, k_y)$, forms a four-dimensional parameter space $(\xi_x, \xi_y, k_x, k_y)$. Because the four parameters are all periodic, the periodic gauge can be imposed on eigenstates, which is $|\psi_n(\xi_x+a, \xi_y, k_x, k_y)\rangle = |\psi_n(\xi_x, \xi_y+a, k_x, k_y)\rangle = |\psi_n(\xi_x, \xi_y, k_x+2\pi a, k_y)\rangle = |\psi_n(k_x, \xi_y, k_x, k_y+2\pi a)\rangle = |\psi_n(\xi_x, \xi_y, k_x, k_y)\rangle$, where $|\psi_n(\xi_x, \xi_y, k_x, k_y)\rangle$ denotes the eigen Bloch state with parameter $(\xi_x, \xi_y, k_x, k_y)$ and band number $n$. If we spatially divide the 2D PnC into two continuous parts and apply different translation vectors to them respectively, two domains will be formed in this PnC, accompanied by the appearance of a 1D domain boundary.

As shown in Fig. 1(c), we divide the 2D PnC into two parts along the $y$-direction and apply translation vectors $\xi_1$ and $\xi_2$ to the left and right halves, respectively. In this way, we construct a domain boundary along the y-direction in the PnC. Since this domain boundary is straight and parallel to the high symmetry direction of the PnC, such a 2D PnC can be equivalently reduced to 1D, and the 1D domain boundary can be equivalently reduced to zero-dimensional (0D). For 1D PCs and PnCs containing different domains, the topological properties of 0D boundaries can be investigated by examining the Zak phases of the domains [36,37]. This method also works for our system. Fig. 1(d) shows the Zak phases of our PnC as a function of $\xi_x$ (keeping $\xi_y$ constant). When $\xi_x$ changes from -0.5a to 0.5a, the Zak phase increases continuously from $-\pi$ to $\pi$ [35]. Note that whatever $\xi_x$ is introduced, the PnC's band structure remains the same.

The Zak phase distinction between two domains guarantees deterministic interface states, proven extensively in electromagnetic [38-44] and mechanical systems [45]. In our PnC, we introduce a homogeneous dislocation [46] to achieve this Zak phase distinction. Specifically, the x (y) components of the translation vector of the two domains are set to negative (consistent), *i.e.*, $\xi_L = (\xi_x, \xi_y)$ and $\xi_R = (-\xi_x, \xi_y)$. Also, to avoid damaging the integrity of the column structure, the value of $\xi_x$ is set between $0.25a$. This method greatly reduces the difficulty of sample preparation (see SI Note 3 [35]). Under this scheme, we successfully constructed the deterministic interface states, as their band structures are shown in Fig. 1(e). The calculated displacement field distribution of one representative interface state is shown in Fig. 1(f), confirming that the elastic energy is mainly localized in the domain boundary, *i.e.*, the homogeneous dislocation.

Note that these interface states are submerged into the bulk when there is no complete bandgap (see SI Note 4 [35]). Moreover, if the y (x) components of the translation vector of the two domains are set to negative (consistent), *i.e.*, $\xi_L = (\xi_x, \xi_y)$ and $\xi_R = (\xi_x, -\xi_y)$, no interface states can be constructed (see SI Fig. S5 [35]).

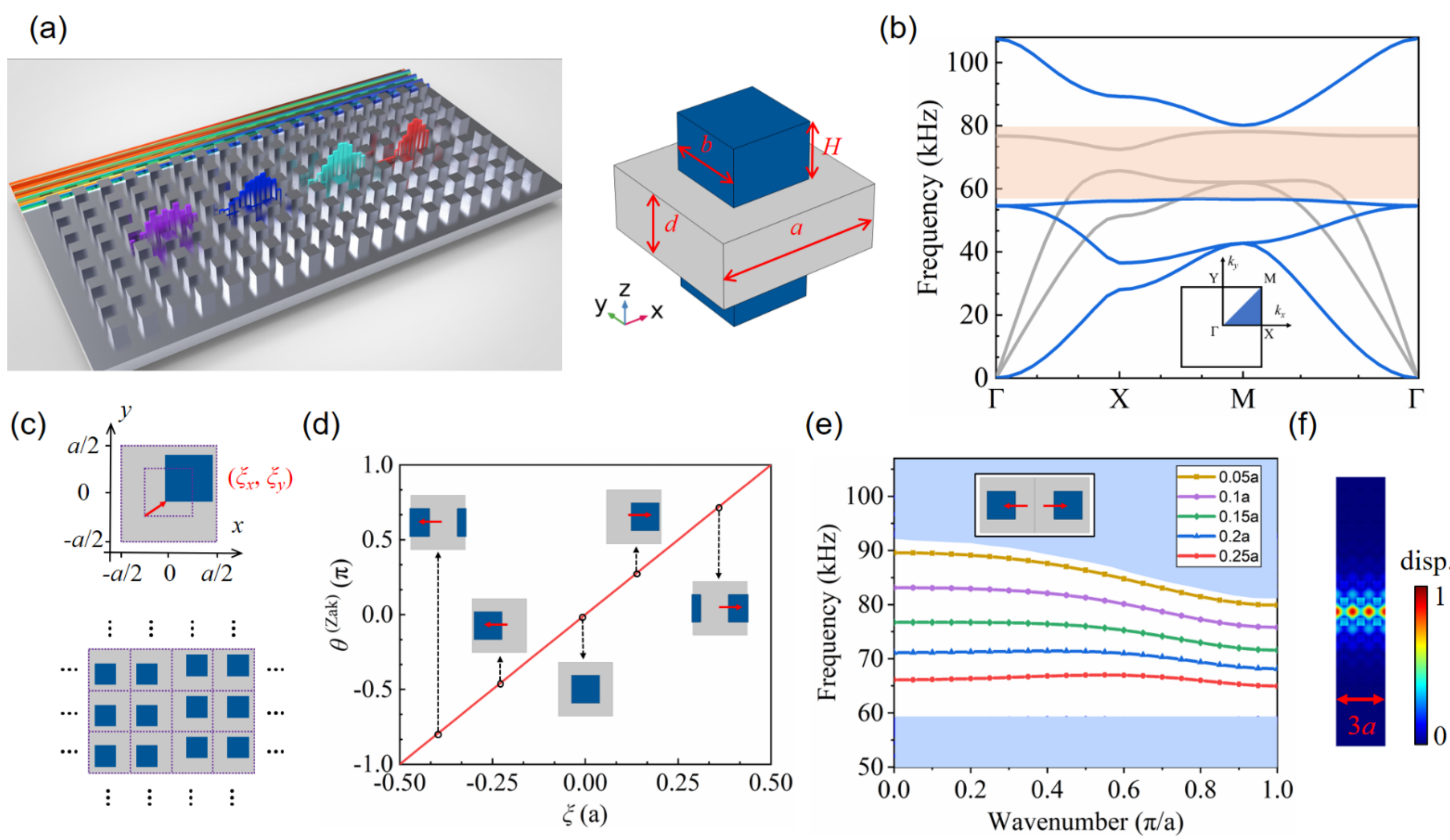

FIG. 1 Elastic deterministic interface states on a homogeneous dislocation. (**a**) Schematic of our 2D elastic rainbow PnC, with identical columns on both sides of a plate. The enlarged view shows its unit cell in a square lattice. (**b**) Calculated 2D band structure. Blue (gray) lines correspond to the out-of-plane flexural modes (the extensional and SH modes). The inset is the Brillouin zone. (**c**) Domain boundary formation by applying a 2D translation vector $\xi$ to different parts of the PnC. (**d**) Zak phase evolution as a function of $\xi_x$. (**e**) Band structures of the deterministic interface states on a homogeneous dislocation. The inset shows a schematic of the dislocation, where $\xi_L = (\xi_x, \xi_y)$ and $\xi_R = (-\xi_x, \xi_y)$. Blue shaded regions are the bulk bands. (**f**) Calculated out-of-plane displacement field distribution of an interface state ($\xi_x$=0.1$a$, $k$=1).

## III. Modulation of the interface states by the translational vector $\xi$

The interface states' operation bandwidth and group velocities can be tuned by the translational vector $\xi$. Fig. 2(a) show the interface states' group velocity spectra under five different $\xi_x$. The group velocities are determined by $V_g = 2\pi \cdot df/dk$, derived from the calculated band structures (Fig.1(e)). All the spectra show a maximum value near the middle of their operation bandwidth. Also, at their frequency extrema (both minima and maxima), all group velocities become zero, owing to the differentiable nature of the phononic bands. To demonstrate the properties of boundary states more intuitively under different $\xi_x$, we also plot a group velocity distribution as a function of continuous $\xi_x$, as shown in Fig. 2(b). Between 65 kHz and 82 kHz (within the complete bandgap of our PnC), as $\xi_x$ increases from 0 to 0.25*a*, both the maximum group velocity and the frequency extrema of the interface states decrease.

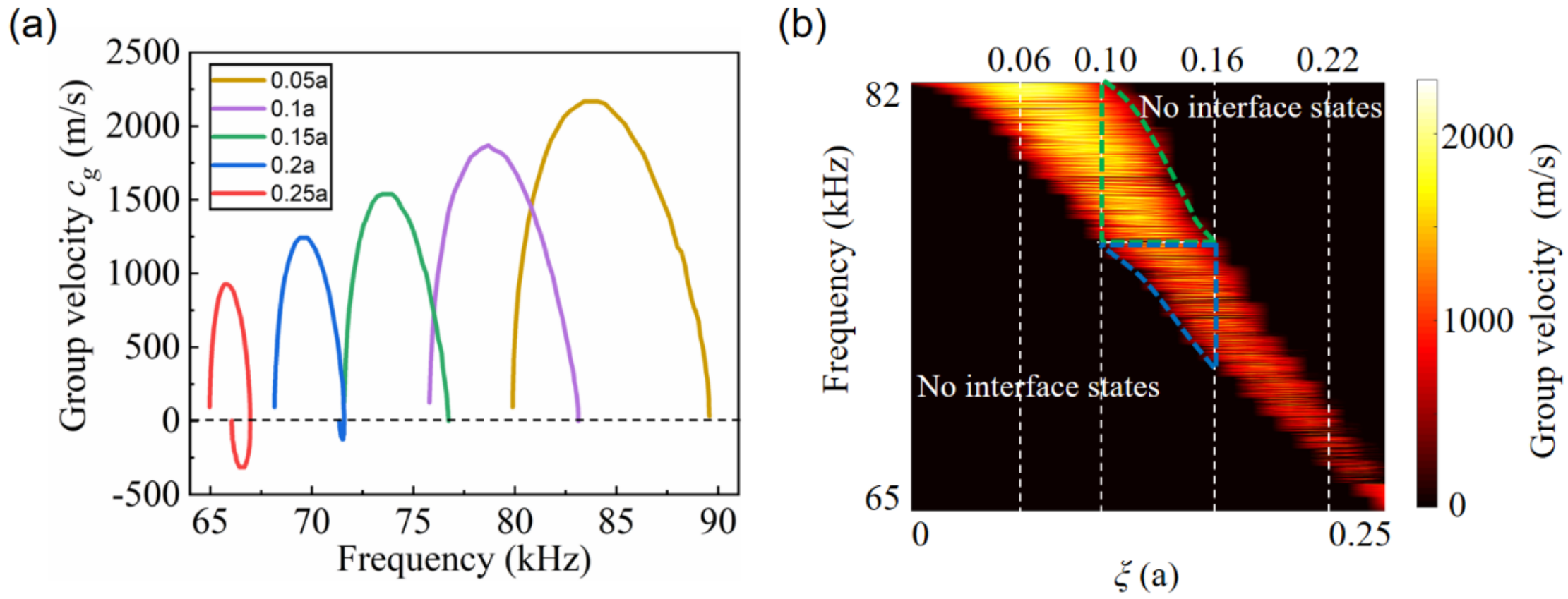

FIG. 2 Group velocity of the interface states. (**a**) Group velocity spectra of the interface states when $\xi_x$ equals to 0.05*a*, 0.1*a*, 0.15*a*, 0.2*a*, 0.25*a*, respectively. (**b**) Group velocity distribution with continuous variation of $\xi_x$. The bright region indicates interface states with a certain operation bandwidth; two dark regions represent the absence of interface states.

## IV. Elastic rainbow trapping through a gradually-tuned boundary

Rainbow trapping is a phenomenon in which states with different frequency components are dispersed and trapped at different spatial locations [22]. As mentioned above, since the frequency extrema of the elastic interface states can be tuned via the geometrical translational vector $\xi$, and the group velocities at the extrema are always zero, these boundaries can therefore be utilized for rainbow trapping of elastic waves.

In the experiment, we fabricated a PnC containing one boundary with gradually-tuned $\xi_x$ from 0.1*a* to 0.16*a*, as the sample shown in Fig. 3(a). A broadband longitudinal transducer, acting as an elastic wave source, was placed at one port of the boundary. A scanning laser vibrometer was used to record the out-of-plane displacement of elastic waves nearby the boundary, working at different frequencies. The experimental scan area is indicated by the blue dashed box in Fig. 3(a), where $\xi_x$ is from 0.105*a* to 0.1525*a*. The results are shown in Fig. 3(b).

When the source was placed at the port where $\xi_x$=0.1*a*, as frequency increases, the distance the elastic waves travel along the boundary (and eventually stop caused by the zero-group velocity) is reduced. In this case, since the near-source boundary ($\xi_x$=0.1*a*) only supports elastic waves with frequencies from ~73.5 kHz to ~82 kHz, only waves in this frequency range can propagate along the boundary. As the wave propagates, $\xi_x$ gets larger; thus, the supported frequency range gets narrower. For the elastic wave in lower frequency, *e.g.,* 73.5 kHz, it can travel through the entire gradually-tuned boundary. By contrast, the elastic wave in higher frequency, e.g., 78.5 kHz, can only travel a short distance before being forced to stop. The experimental results and analysis match the calculation in Fig. 2(b), which frames out rainbow trapping frequencies (see the green region in the figure) for this scenario.

The rainbow trapping would be different when the source was placed at the port where $\xi_x$=0.16$a$. In this case, since the near-source boundary ($\xi_x$=0.16$a$) only support elastic wave with frequencies from ~70 kHz to ~73.5 kHz, only waves in this frequency range can propagate along the boundary. Only the elastic wave in higher frequency, *e.g.,* 73.2 kHz, can travel through the entire gradually-tuned boundary. Simulation is performed for this scenario, as shown in Fig. 3(c), matching well with the calculation in Fig. 2(b) (see the blue region in the figure).

In both scenarios, when elastic waves of different frequencies are dispersed in the gradually-tuned boundary, they will resonant where they stop, simply owning to their group velocity being reduced to zero at the location. Additionally, this elastic rainbow trapping applies not only to the flexural modes we demonstrate here but also to the in-plane modes. Within the complete in-plane modes bandgap, as shown in Fig. 1(c), our simulation validates that the in-plane modes can also be rainbow trapped by taking a similar approach (see SI Note 6 [35]).

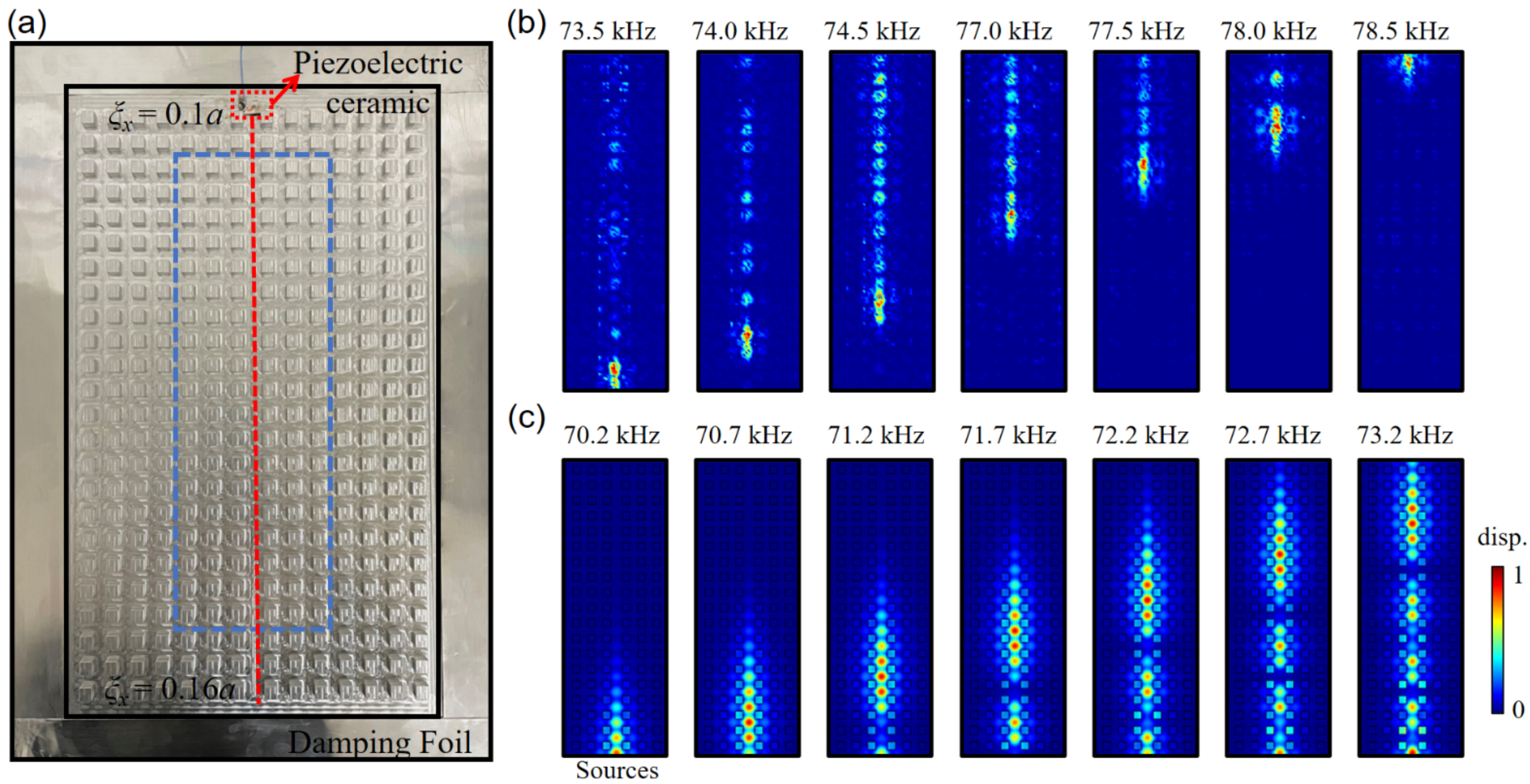

**FIG. 3** Elastic rainbow trapping. (**a**) Photograph of our elastic PnC with a gradually-tuned boundary. The $\xi_x$ equals 0.1$a$ and 0.16$a$ on the top and bottom ports of the boundary, respectively. The PnC has 24 layers along the boundary direction, and the difference of $\xi_x$ between each adjacent layer is 0.0025$a$. (**b**) and (**c**) Out-of-plane displacement field distributions in the PnC, when elastic waves of different frequencies enter the boundary from the top port (experimental results) and bottom port (simulated results), respectively.

### V. Elastic rainbow concentrator utilizing the gradually-tuned boundary

As recently proposed [33], a rainbow concentrator can collect weak signals multi-directionally and assign different frequency components to different spatial locations. Compared with traditional concentrators, *e.g.*, in plasmonics [47] or transformation optics [48], which either operate in a narrow bandwidth or have no frequency resolution, a rainbow concentrator operates at multiple frequencies and therefore greatly improves the performance of a single device.

In the rainbow trapping experiment, if we construct a boundary with $\xi_x$ ranging from $\xi_1$ to $\xi_2$ ($\xi_1 < \xi_2$), elastic resonances of frequencies ranging from $f_{\xi 2}$ to $f_{\xi 1}$ may all be supported in the boundary but at different locations, where $f_{\xi 2}$ ($f_{\xi 1}$) is the minimum (maximum) frequency of the boundary states in the case of $\xi_2$ ($\xi_1$). Since the elastic waves need to enter the boundary through the two ports (either one), this forces all $\xi$ in the interval from the entrance port to the resonant position to support the resonant frequency (at the resonant position). However, if the elastic waves do not need to reach the resonant positions along the boundary, the operation bandwidth of the whole device will be greatly broadened. This idea also inspired us to realize an elastic rainbow concentrator/resonator.

Our proposed elastic rainbow concentrator is demonstrated using another PnC, containing one boundary with a more significant gradually-tuned $\xi_x$ from $0.06a$ to $0.22a$, as the sample shown in Fig. 4(a). In this PnC, there are only four layers of the columns on either side of the boundary. In the boundary direction, there are 33 layers, and the $\Delta\xi_x$ between each adjacent layer is $0.005a$. Eigenvalue calculation in Fig. 4(b) shows that 28 resonant states exist in this PnC, which are continuously distributed at different boundary positions with a monotonic variation of frequency.

In this configuration, since the PnC is not thick enough to fully support its bandgap, rainbow resonant states can be excited by elastic waves that do not propagate along the boundary. Experimentally, a row of broadband longitudinal transducers, analogous to a plane wave source for the elastic waves, was placed on the side of the PnC. Thus, the plane elastic waves pumped by the transducers are incident on the PnC perpendicular to the boundary direction. The scanning laser vibrometer records how the resonant modes are excited at different frequencies. Fig. 4(c) shows the measured intensity spectrum of these resonant states when excited. Their resonant frequencies are in perfect agreement with the results of the eigenvalue calculations.

Displacement field distributions of the resonant states at different frequencies are measured, as shown in Fig. 4(d) (all resonant states see SI Fig. S8 [35]). The experimental results correspond well with the simulation ones (shown in Fig. 4(e)), validating the elastic rainbow concentrating. Note that these resonant states' mode volumes are relatively large due to the small frequency separation between adjacent resonant modes in this PnC, i.e., less than 0.5kHz (about 0.6% relative bandwidth). Each resonance appears as a long grain with multiple wavelengths along the boundary. Although they are not excited by elastic waves propagating along the boundary, they still exhibit asymmetric energy distribution in the boundary direction. Each resonant state is shaped like a tadpole, exhibiting the strongest energy localization on the side with larger $\xi_x$.

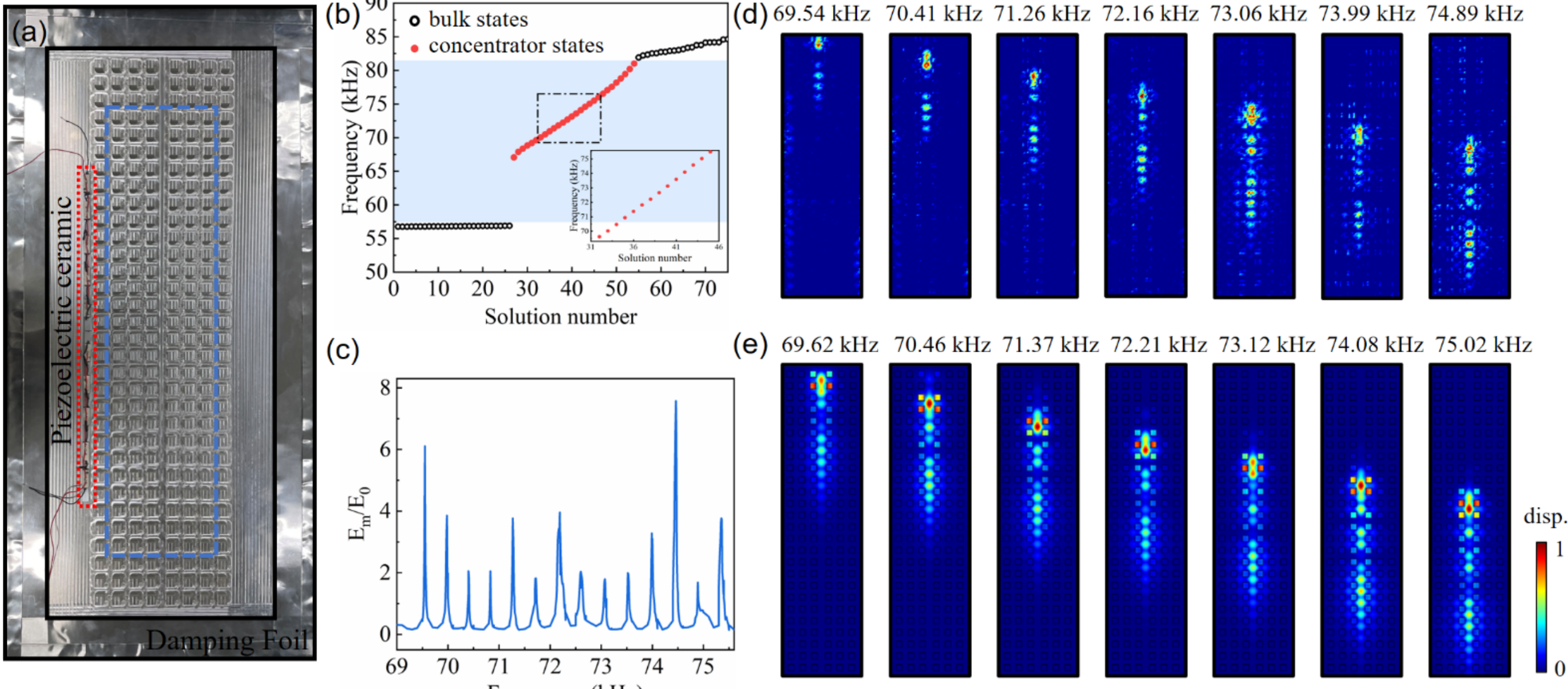

**FIG. 4** Elastic rainbow concentrator. (**a**) Photograph of another elastic PnC with a gradually-tuned boundary. Here, $\xi_x$ equals from 0.06$a$ (at the bottom) to 0.22$a$ (at the top). Only four layers of columns are assembled in the PnC on both sides of the boundary. (**b**) Eigenvalue calculation of the whole PnC. Frequency-separated (red) dots represent the elastic resonant modes, appearing at different positions along the boundary. (**c**) Normalized elastic intensity spectrum of these resonant modes when excited. $E_m$ and $E_0$ represent the maximal intensity measured in the whole PnC and the intensity measured near the plane wave source, respectively. (**d**) and (**e**) Measured and simulated out-of-plane displacement field distributions of the rainbow resonant modes at different frequencies.

### VI. Diverse design for the elastic rainbow concentrator

In our elastic rainbow concentrator, the number of the resonant states in the boundary is determined by the number of different $\xi_x$. If the value range of $\xi_x$ is determined, *e.g.,* from 0.06$a$ to 0.22$a$ in our last experiment, the smaller the $\Delta\xi_x$ between each adjacent layer will result in more resonant states in the same fixed frequency range.

Figs. 5(a) and 5(b) show two PnC similar to the one in Fig. 4(a), but there are fewer layers in the boundary direction. The $\Delta\xi_x$ between each adjacent layer in the two samples is 0.01$a$ and 0.02$a$, respectively. According to eigenvalue calculations shown in Figs. 5(c) and 5(d), only 13 and 7 rainbow resonant states exist in the two PnC, respectively. The number of these states is linearly related to the number of layers along the boundary.

Displacement field distributions of the resonant states in the two cases are shown in the insets of the figures. As the frequency separation between the resonant states increases, their mode volumes decrease approximately linearly. In the PnC of Fig. 5(a) (Fig. 5(c)), the frequency separation between adjacent resonant states is about 1.76 kHz (0.82 kHz), and their modes volume is only about 1/2 (1/4) of the one in Fig. 4(a).

Clearly, by designing different ranges and gradients of $\xi_x$ for the boundary of the PnC, the frequency range, number, and mode volume of the rainbow resonant states can be effectively tuned. It provides a high degree of freedom for elastic wave manipulation and is promising for applications, *e.g.,* acoustofluidics, elastic energy harvesting, and spatial wave switch [13, 21].

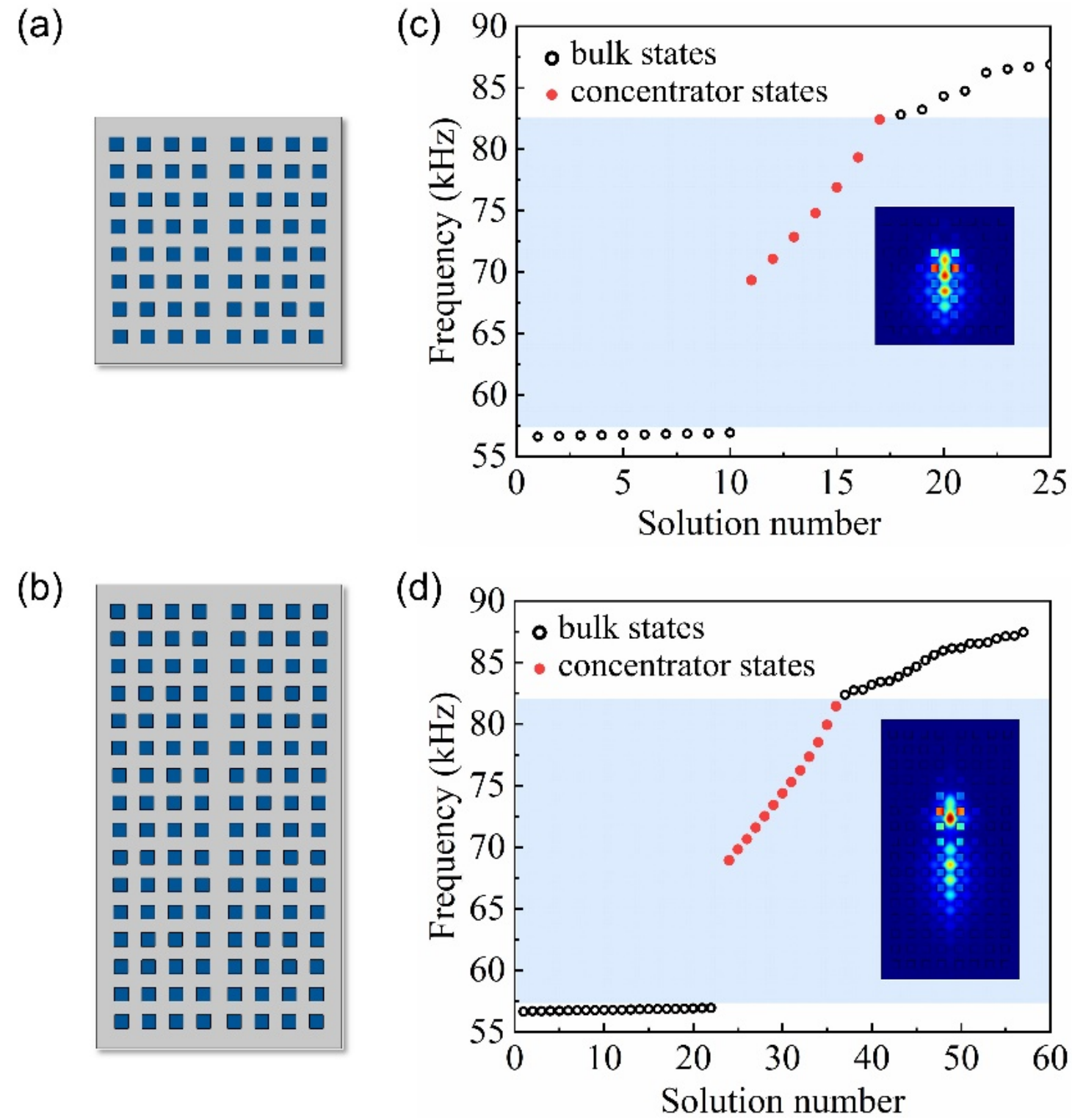

**FIG. 5** Configurable rainbow concentrators. (**a**) and (**b**) PnCs similar to the one in Fig. 4(a) but with fewer layers along the gradually-tuned boundary. $\Delta\xi$ between each adjacent layer in the two samples is 0.02$a$ and 0.01$a$, respectively. (**c**) and (**d**) eigenvalue calculations of the two PnCs. The insets show simulated out-of-plane displacement field distributions of the rainbow resonant states, whose operating frequencies are 74.8 kHz and 73.43 kHz, respectively.

## VII. Conclusions

This paper verifies a recently proposed exotic wave localization experimentally, *i.e.*, rainbow concentrating, exemplarily on an elastic system. We demonstrate that by bringing homogeneous dislocation, *i.e.*, a line defect, into a plate PnC, two separated PnCs will have distinct Zak phases due to different directional translations, leading to deterministic interface states along with the defect. Different translations can modulate the group velocity of these interface states. Thus, we could delicate rainbow trapping and rainbow concentrating for elastic waves. Compared with previous theoretical approaches, the elastic rainbow effect realized by our present method is simple and configurable, not only for design but also for practical sample implementation and miniaturization. This work undoubtedly provides diversity for the precise manipulation of elastic waves in both frequency and spatial domains. It is promising for applications, e.g., nondestructive evaluation [49], wideband energy harvesting [13], and spatial wave switch [21]. Moreover. this method of modulating the boundary states by graded translational deformations can be extended to high-frequency microwave-acoustics, e.g., surface acoustic waves (SAWs), for on-demand storage and extraction of chip-scale phonons used in acoustofluidics [50] and phononic computing [51] and information processing [52].

## ACKNOWLEDGMENTS

The work was jointly supported by the National Key R&D Program of China (Grants No. 2021YFB3801801, No. 2017YFA0305100, and No. 2017YFA0303702) and the National Natural Science Foundation of China (Grants No. 11890702, No. 92163133, and No. 51732006). We also acknowledge the support of the Fundamental Research Funds for Central Universities.

## APPENDIX A: Methods

**Numerical simulations.** We performed a 3D finite element simulation using the commercial software of COMSOL Multiphysics with the Structural Mechanics Module. The aluminum alloy plate density, Young's modulus, and Poisson's ratio were 2810 kg $m^{-3}$, 70.368 GPa, and 0.314, respectively. For the finite element simulation, meshes are set to $<a/10$ to ensure the accuracy of the calculation. An Eigenfrequency Study with Parameter Sweep calculated the PnC's band structure. The Frequency Domain Study obtains displacement field distributions of the elastic waves.

**Sample preparation.** Using a precision CNC milling machine, we prepared the PnC samples on aluminum alloy plates with a fixed thickness of ~2.2 cm. Young's modulus and Poisson's ratio of the plates were determined by adopting the ultrasonic scattering echo method. A 2-inch margin on the plate boundary was left for 3M™ Damping Foil 2552, to reduce unwanted mechanical reflection from the PnC edges during the experiments.

**Experimental measurements.** Broadband piezoelectric transducers (centered at 75 kHz) were used as actuators, bonded to the surface of the PnC with 3M™ Scotch-Weld™ Structural Acrylic Adhesives. An arbitrary wave generator generates alternating signals of tens of kHz, amplified to excite the transducer. A laser Doppler vibrometer was used to record the out-of-plane displacement of the elastic waves, offering a 2D distribution field after scanning.